\documentclass[letter]{aa}  

\usepackage{graphicx}
\usepackage{txfonts}
\usepackage{natbib}
\bibpunct{(}{)}{;}{a}{}{,}
\begin{document}

   \title{The effect of variations in magnetic field direction from turbulence on kinetic-scale instabilities}

\author{Simon Opie\inst{1}\and Daniel Verscharen\inst{1} \and Christopher H. K. Chen\inst{2}\and Christopher J.~Owen\inst{1} \and Philip A.~Isenberg\inst{3}}

\institute{Mullard Space Science Laboratory, University College London, Holmbury St Mary, Dorking, Surrey RH5~6NT, UK \\ \email{simon.opie.18@ucl.ac.uk} \and Department of Physics and Astronomy, Queen Mary University of London, Mile End Road, London E1~4NS, UK \and Space Science Center, University of New Hampshire, 105 Main Street, Durham NH 03824, USA}


 
  \abstract {At kinetic scales in the solar wind, instabilities transfer energy from particles to fluctuations in the electromagnetic fields while restoring plasma conditions towards thermodynamic equilibrium. We investigate the interplay between background turbulent fluctuations at the small-scale end of the inertial range and kinetic instabilities acting to reduce proton temperature anisotropy. We analyse in-situ solar wind observations from the Solar Orbiter mission to develop a measure for variability in the magnetic field direction. We find that non-equilibrium conditions sufficient to cause micro-instabilities in the plasma coincide with elevated levels of variability. We show that our measure for the fluctuations in the magnetic field is non-ergodic in regions unstable to the growth of temperature anisotropy-driven instabilities. We conclude that the competition between the action of the turbulence and the instabilities plays a significant role in the regulation of the proton-scale energetics of the solar wind. This competition depends not only on the variability of the magnetic field but also on the spatial persistence of the plasma in non-equilibrium conditions.}
  
   
   
   
   

  \keywords{solar wind --
                turbulence --
                kinetic instabilities}
               
    \titlerunning{Effect of magnetic field variations on kinetic instabilities} 
    \authorrunning{S. Opie et al.} 
   \maketitle 
   

\section{Introduction}

The solar wind is a nearly collisionless plasma and as such exhibits non-equilibrium conditions that lead to the creation of micro-instabilities  \citep{matteini_ion_2012,alexandrova_solar_2013,klein_majority_2018}. Linear and quasilinear Vlasov--Maxwell theory predicts that kinetic-scale instabilities driven by temperature anisotropy with respect to the magnetic field restore the plasma towards thermal equilibrium  \citep{hollweg_new_1970,gary_proton_1976,gary_theory_1993}. These theoretical descriptions often assume a constant background on which the unstable fluctuations are added as a perturbation. However, the real, turbulent solar wind does not provide such a constant background, with the presence of inhomogeneities across the spatial and temporal scales over which the instabilities are predicted to act \citep{bruno_solar_2013,matthaeus_nonlinear_2014,verscharen_multi-scale_2019}. 

The effective action of proton temperature anisotropy-driven instabilities in the solar wind is often inferred in the literature from comparisons of the distribution of observed data and its constraints in the $T_{\perp}/T_{\parallel}$-$\beta_{\parallel}$ parameter space, where 
\begin{equation}\label{beta}
\beta_{\parallel}\equiv \frac{8\pi nk_{\mathrm B}T_{\parallel}}{B^2},
\end{equation}
$B$ is the magnitude of the magnetic field, $n$ is the proton number density, $k_{\mathrm B}$ is the Boltzmann constant,  $T_{\perp}$ is the proton temperature perpendicular to the magnetic field, and $T_{\parallel}$ is the proton temperature parallel to the magnetic field \citep[e.g.,][]{marsch_temperature_2004,hellinger_solar_2006, bale_magnetic_2009,chen_multi-species_2016,opie_conditions_2022}. The thresholds of the anisotropy-driven instabilities set limits to the distribution of the data in $T_{\perp}/T_{\parallel}$--$\beta_{\parallel}$ parameter space \citep{gary_mirror_1992,gary_proton_2001,kasper_windswe_2002}. 

A common analytical approximation for the thresholds of the anisotropy-driven instabilities is given in the parametric form
    \begin{equation} \label{ethresh}
        \frac{T_{\perp}}{T_{\parallel}} = 1 + \frac{a}{\left(\beta_{\parallel} - c\right)^b},
    \end{equation}
where $a$, $b$, and $c$ are fit parameters with values specific to each instability and to a given maximum growth rate $\gamma_m$ \citep{hellinger_solar_2006}.
The oblique firehose and the mirror-mode instabilities, which we consider here, approximately provide  outer boundaries to the distribution of stable data both in observations \citep{hellinger_solar_2006,gary_short-wavelength_2015} and in simulations \citep{servidio_proton_2014,hellinger_plasma_2015,riquelme_particle--cell_2015}. 

Solar wind turbulence is mostly non-compressive with a minor component of compressive fluctuations that contribute a relative magnetic energy $\left(\delta |\vec B|/B_0\right)^2$ of a few percent to the turbulent cascade \citep{chen_recent_2016}. Turbulent dissipation of energy is a candidate mechanism to explain the observed anisotropic heating of the solar wind \citep{isenberg_resonant_1984, marsch_turbulence_1991, cranmer_self-consistent_2007, maruca_what_2011, howes_dynamical_2015}. In the context of the expanding solar wind, local heating  and the response of the solar wind to the turbulent fluctuations create non-equilibrium features that displace the plasma into unstable regions of the $T_{\perp}/T_{\parallel}$--$\beta_{\parallel}$ parameter space, beyond the threshold of the instabilities \citep{matteini_parallel_2006,schekochihin_nonlinear_2008,matteini_ion_2012,bott_adaptive_2021}. However, it is unclear how instabilities and turbulence interact at kinetic scales. 

Kinetic plasma simulations show that instabilities can regulate the thermal energetics of the plasma, and that turbulence in the expanding solar wind can both raise and lower anisotropy measured with respect to the magnetic field \citep{matteini_parallel_2006, hellinger_oblique_2008,kunz_firehose_2014,hellinger_plasma_2015,riquelme_particle--cell_2015, hellinger_mirror_2017,markovskii_mirror_2020,bott_adaptive_2021, markovskii_effect_2022}. The oblique firehose instability produces Alfv\'enic modes with zero frequency, linear polarisation, and finite compressivity ($\delta n\ne 0$ and $\delta |\vec B|\ne 0$), and so creates both compressive and non-compressive fluctuations at ion scales \citep{hellinger_new_2000,hellinger_oblique_2008}. Observations and simulations show that the mirror-mode instability generates compressive fluctuations on kinetic scales \citep{bale_magnetic_2009, hellinger_mirror_2017}. Therefore both compressive and non-compressive kinetic-scale fluctuations can be attributed  to the instabilities themselves or to cascaded background turbulence at these scales (or, in fact, a combination of both), and consequently caution must be exercised in their interpretation \citep{bale_magnetic_2009,chandran_constraining_2009, salem_identification_2012, chen_nature_2013, gary_short-wavelength_2015}.  

In this work, we examine non-compressive fluctuations in the magnetic-field direction at scales corresponding to the small-scale end of the inertial range of the turbulence  \citep{kolmogorov_dissipation_1941,tu_mhd_1995}. We assume that these fluctuations predominantly represent local Alfv\'enic fluctuations. By combining  magnetic-field measurements with measurements of the proton parameters, we investigate the action of the oblique firehose and mirror-mode instabilities in this turbulent background.

\section{Data analysis}
\label{data}

\subsection{The magnetic-field variability measure $\sigma_B$}
\label{sigma}

We develop a measure, $\sigma_{B}$, for the directional variability of the magnetic field $\vec B$ using Solar Orbiter data. We use the $8\,\mathrm{vectors/s}$ magnetic-field data from the magnetometer \citep[MAG;][]{horbury_solar_2020} in conjunction with $\approx 10^6$ datapoints at a cadence of  $4\,\mathrm s$ from the Proton-Alpha Sensor (PAS) of the Solar Wind Analyser \citep[SWA;][]{owen_solar_2020}. These data are coincident with the dataset presented by \citet{opie_conditions_2022} and represent predominantly slow solar wind. We do not identify or remove structures such as shocks, coronal mass ejections, or current sheets in the dataset which is taken over 8 separate periods totalling 53 days at an average heliocentric distance of $\sim 0.85\,\mathrm{au}$.

We first derive the magnetic-field unit vector $\vec b=\vec B/|\vec B|$  for each measurement vector $\vec B$ in RTN coordinates. PAS derives the proton moments based on a sampling of 1\,s duration, every 4\,s. We define the centre of the PAS sampling interval as the time $t_{i}$. We associate all $\vec b$ measurements in the interval $[t_i-2\,\mathrm s,t_i+2\,\mathrm s]$ with the PAS interval at time $t_i$. We calculate the standard deviation of the unit-vector component $b_j$ for time interval $t_i$ as
\begin{equation}
    \sigma_{B_j}(t_i) = \sqrt{\frac{\sum\left(b_j - \langle{b_j}\rangle\right)^2}{31}},
    \label{SDstepone}
\end{equation}
where the sum is taken over all 32 magnetic-field measurements associated with the PAS measurement at $t_i$, $\langle\cdot\rangle$ is the average over this time interval of 4\,s duration, and the index $j=(R,T,N)$ marks the field component in RTN coordinates. We then combine the components to
\begin{equation}
    \sigma_B(t_i) = \sqrt{\sigma_{B_R}^2 + \sigma_{B_T}^2 + \sigma_{B_N}^2}.
    \label{SDsteptwo}
\end{equation}
The quantity $\sigma_B$ is a measure of the  variability of the magnetic-field direction (i.e., excluding changes in magnitude) at the 4\,s scale for each combined interval in our SWA/PAS dataset. 
The mean solar wind bulk velocity for our dataset is $427 \,\mathrm{km\,s^{-1}}$. Therefore, the 4\,s temporal scale represents a convected spatial scale of $\sim 1700\, \mathrm{km}$. The mean gyroradius for our dataset is $51.5\, \mathrm{km}$. Thus, $\sigma_B$ represents fluctuations at the small-scale end of the inertial range, in the transition region approaching ion scales \citep{kiyani_dissipation_2015}.

   \begin{figure}
    \centering
    \includegraphics[width=\hsize]{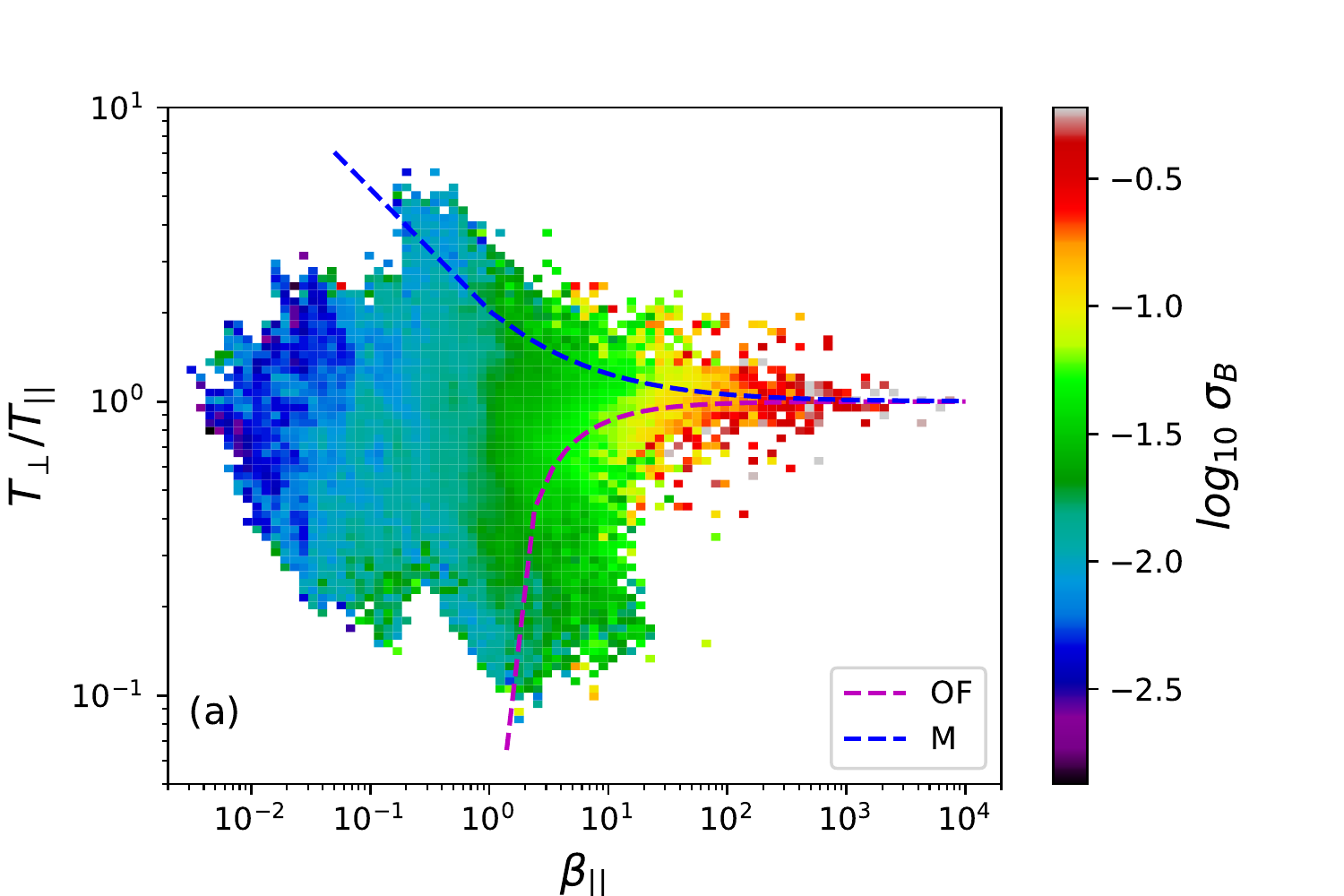}
    \includegraphics[width=\hsize]{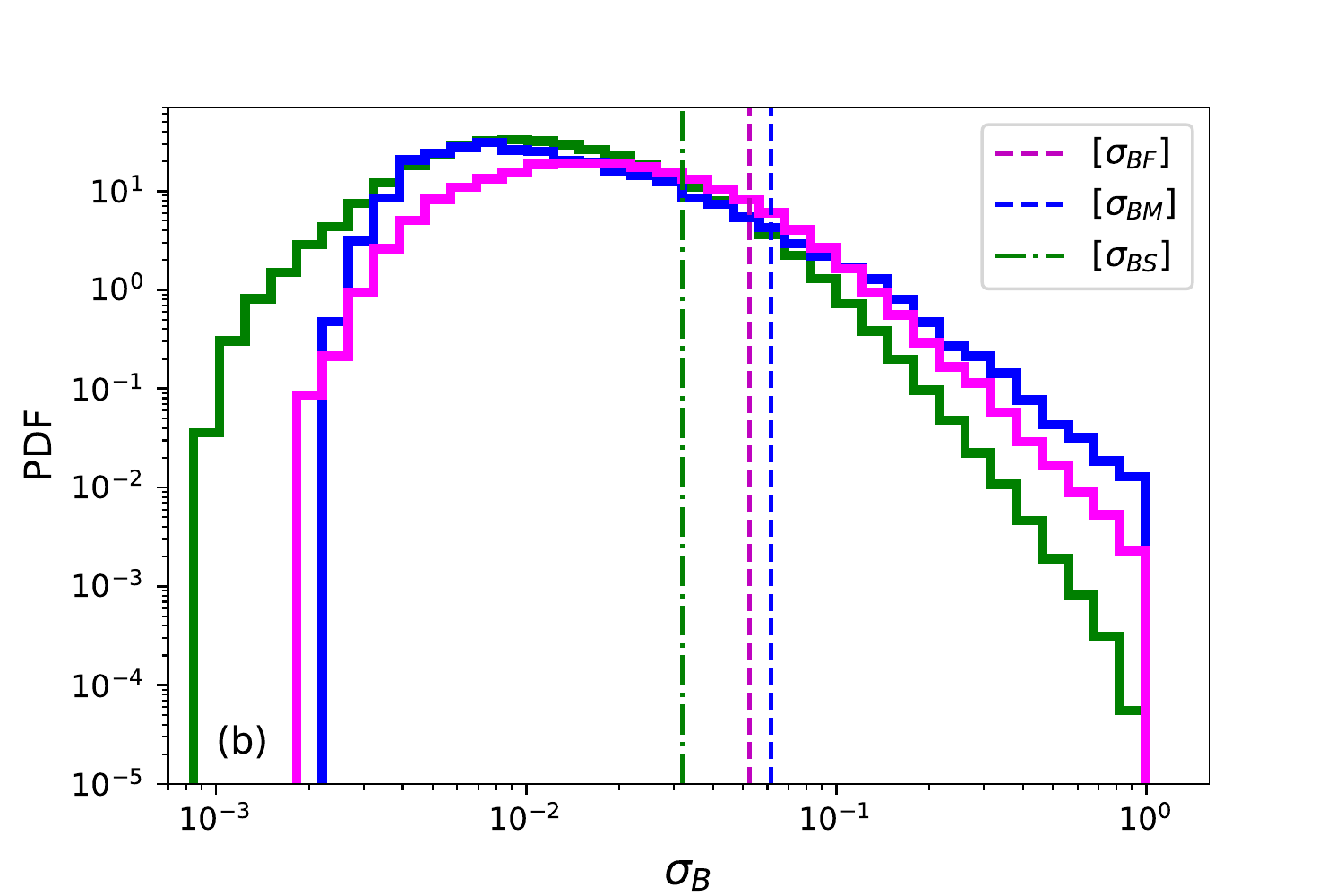}      

\caption{Observed data for $\sigma_B$ plotted as: (a) Distribution of $\sigma_B$ binned and averaged by bincount in $T_{\perp}/T_{\parallel}$--$\beta_{\parallel}$ parameter space. The instability thresholds for the oblique firehose (OF) and mirror-mode (M) instabilities are shown as labelled. (b) PDFs of $\sigma_B$ for oblique firehose unstable (magenta), mirror-mode unstable (blue), and stable (green) points in our dataset. The vertical lines denote the ensemble mean $[\cdot]$ of each dataset. 
\label{fig1draft}}
   
  \end{figure} 
   
   \begin{figure}
   \centering
    \includegraphics[width=\hsize]{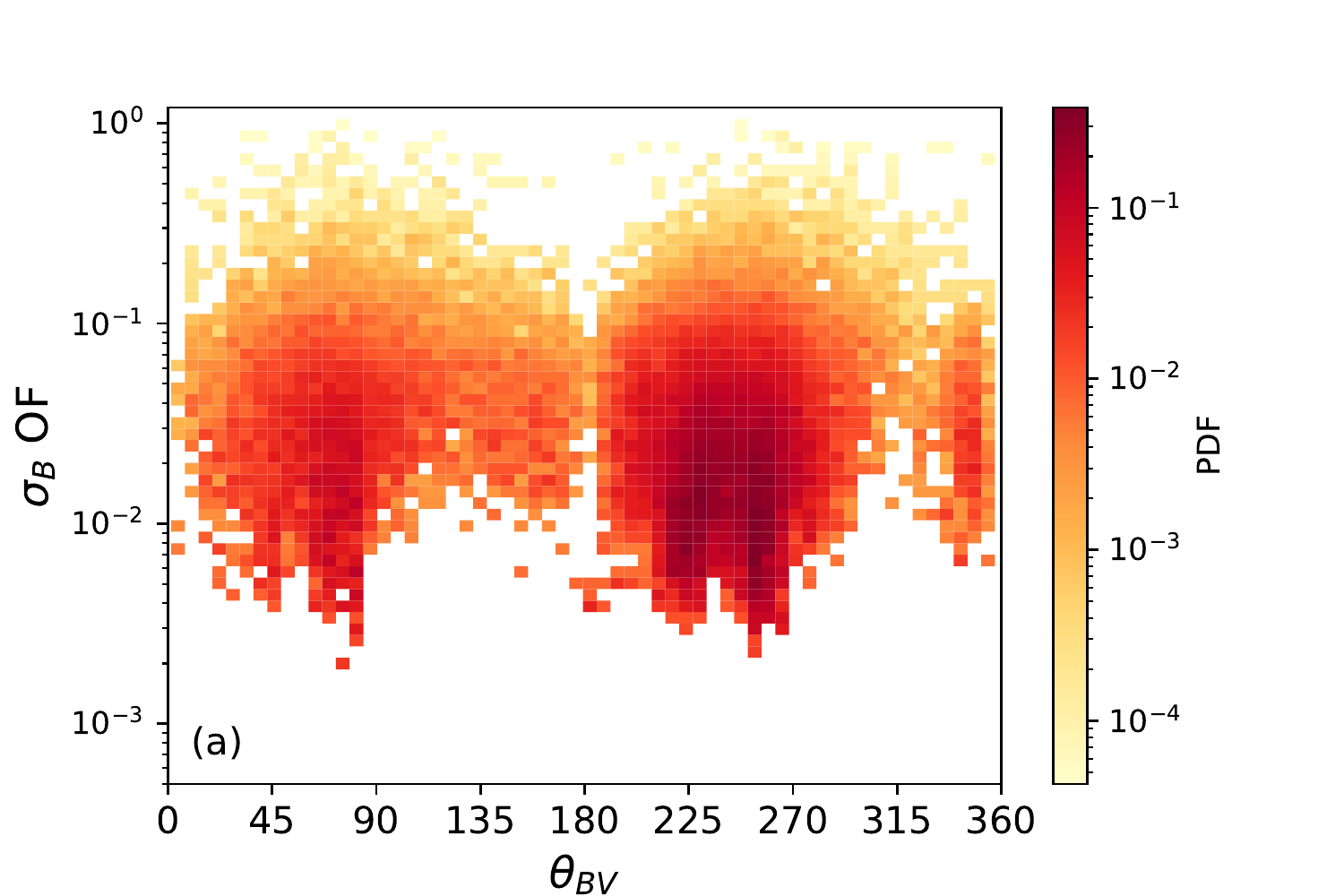}
    \includegraphics[width=\hsize]{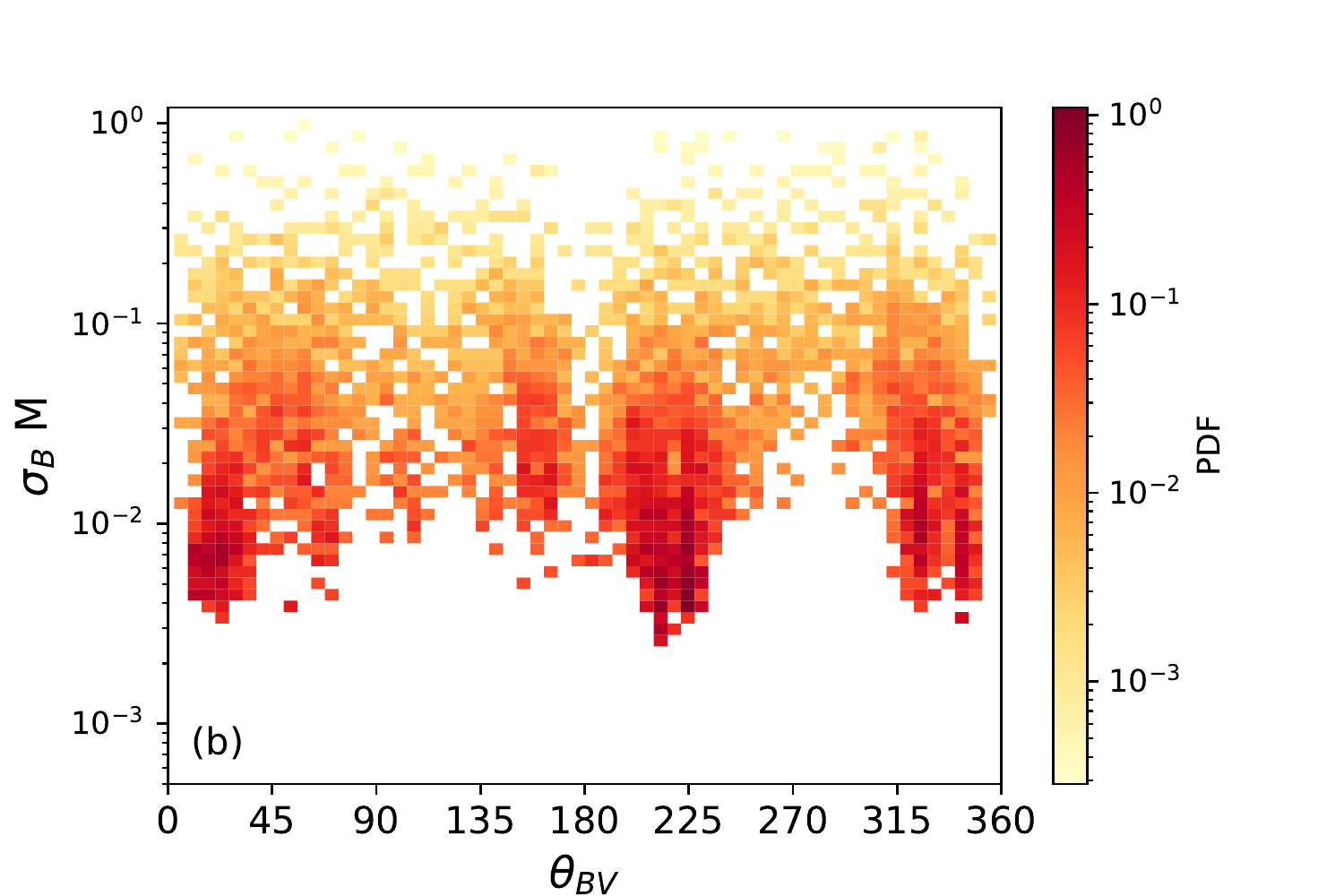}
    \includegraphics[width=\hsize]{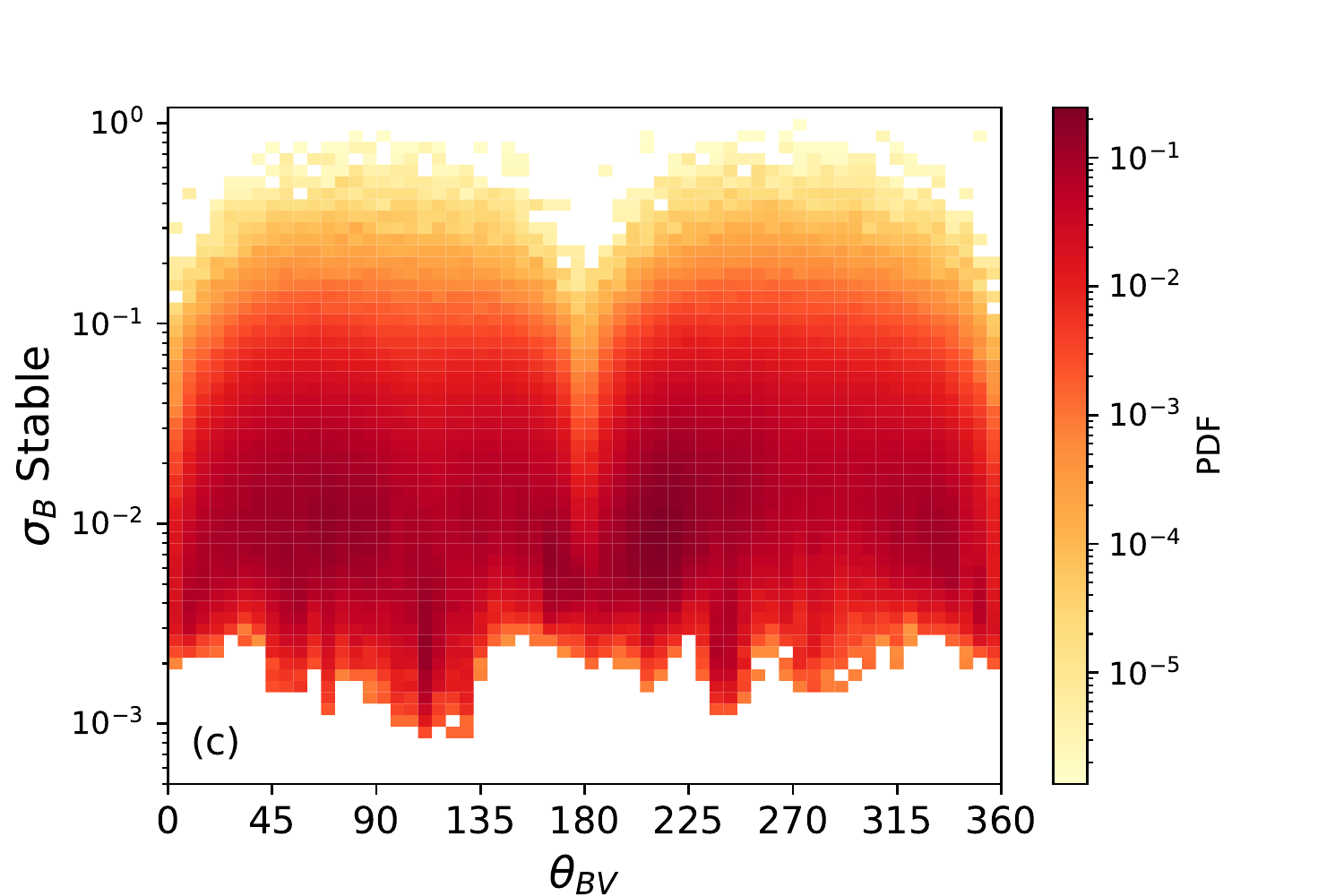}
   \caption{PDF of data in $\sigma_B$--$\theta_{BV}$ parameter space for (a) oblique firehose unstable, (b) mirror-mode unstable, and (c) stable, datapoints.}
              \label{fig2draft}
    \end{figure}

\begin{figure}
   \centering
    \includegraphics[width=\hsize]{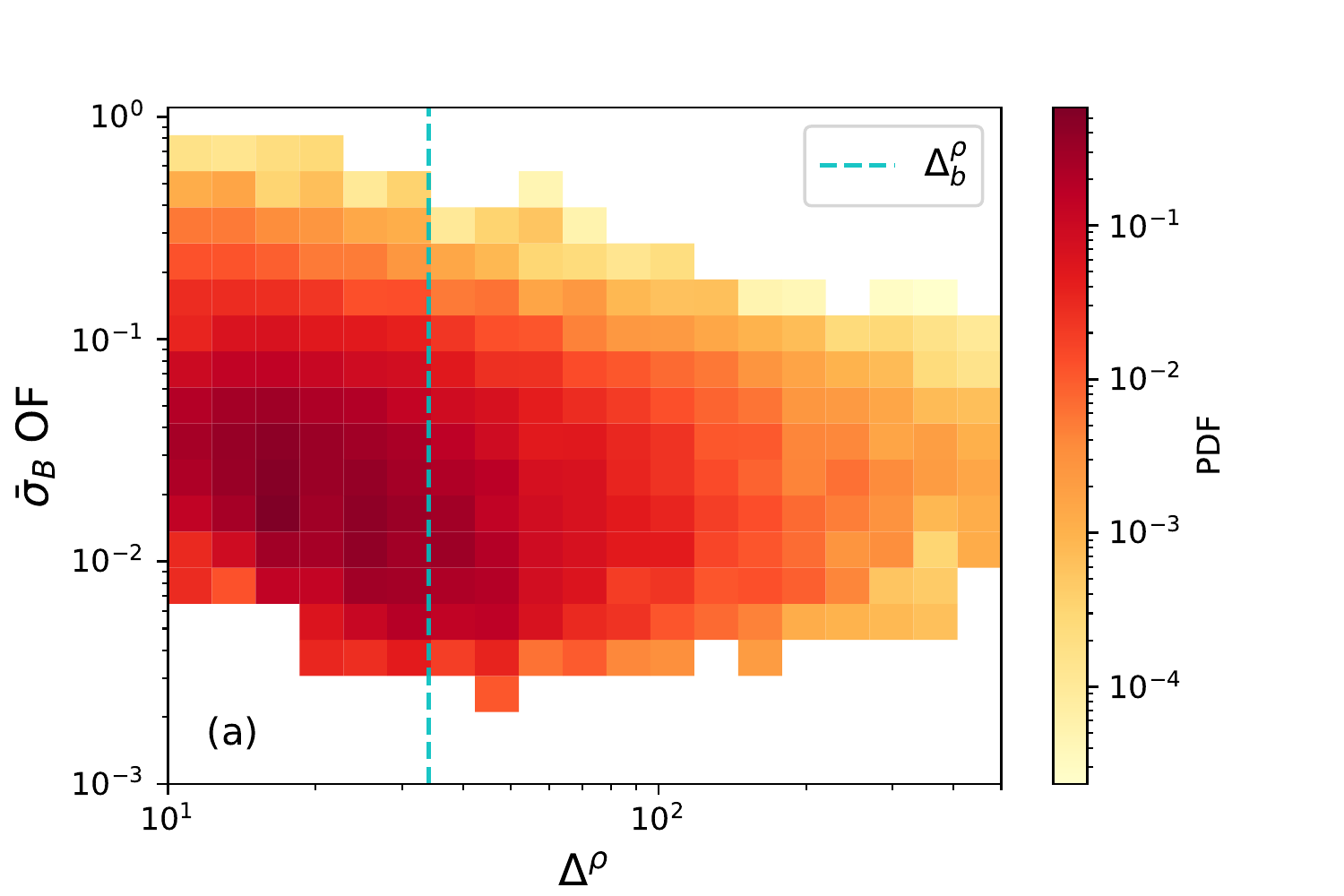}
    \includegraphics[width=\hsize]{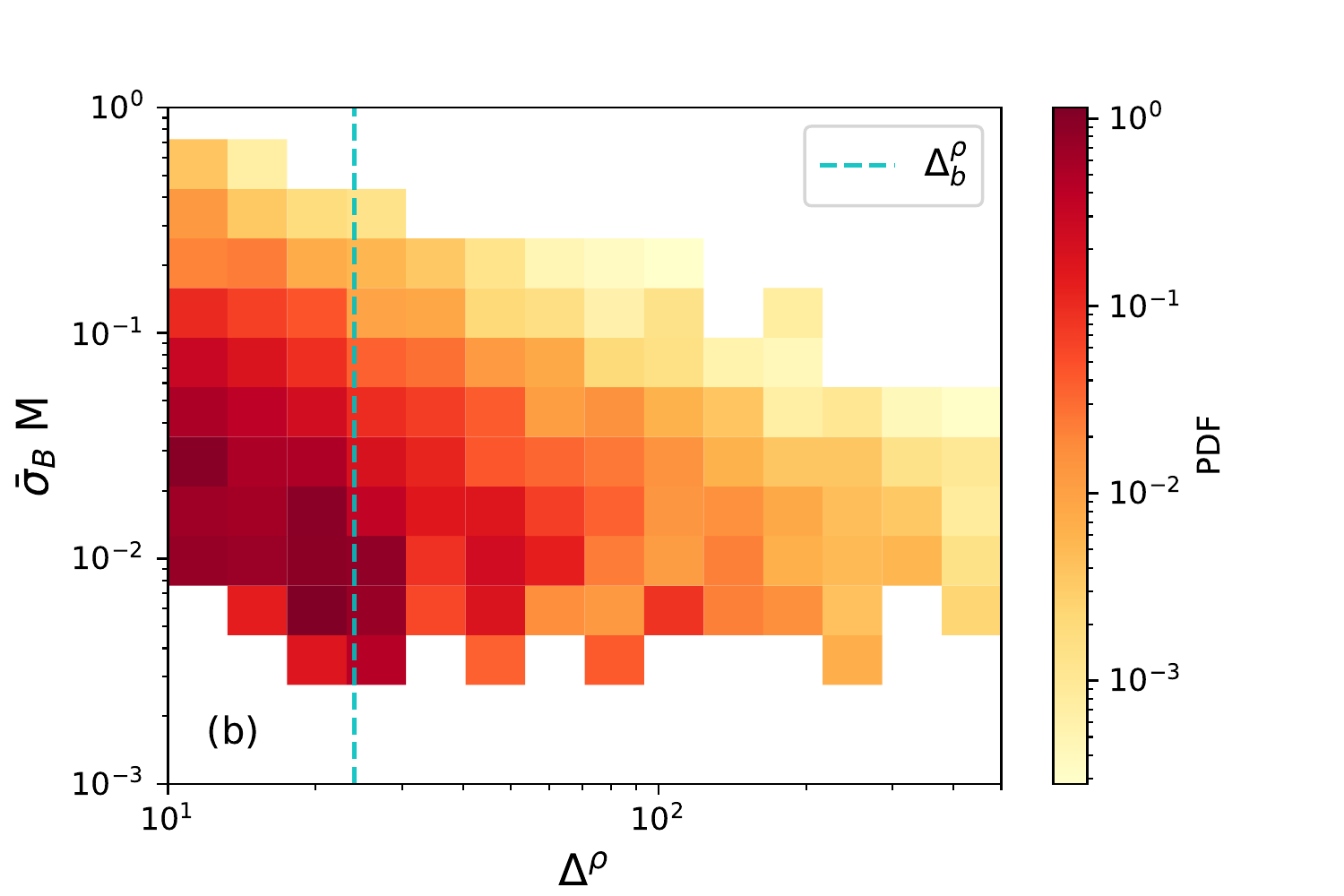}     
    \includegraphics[width=\hsize]{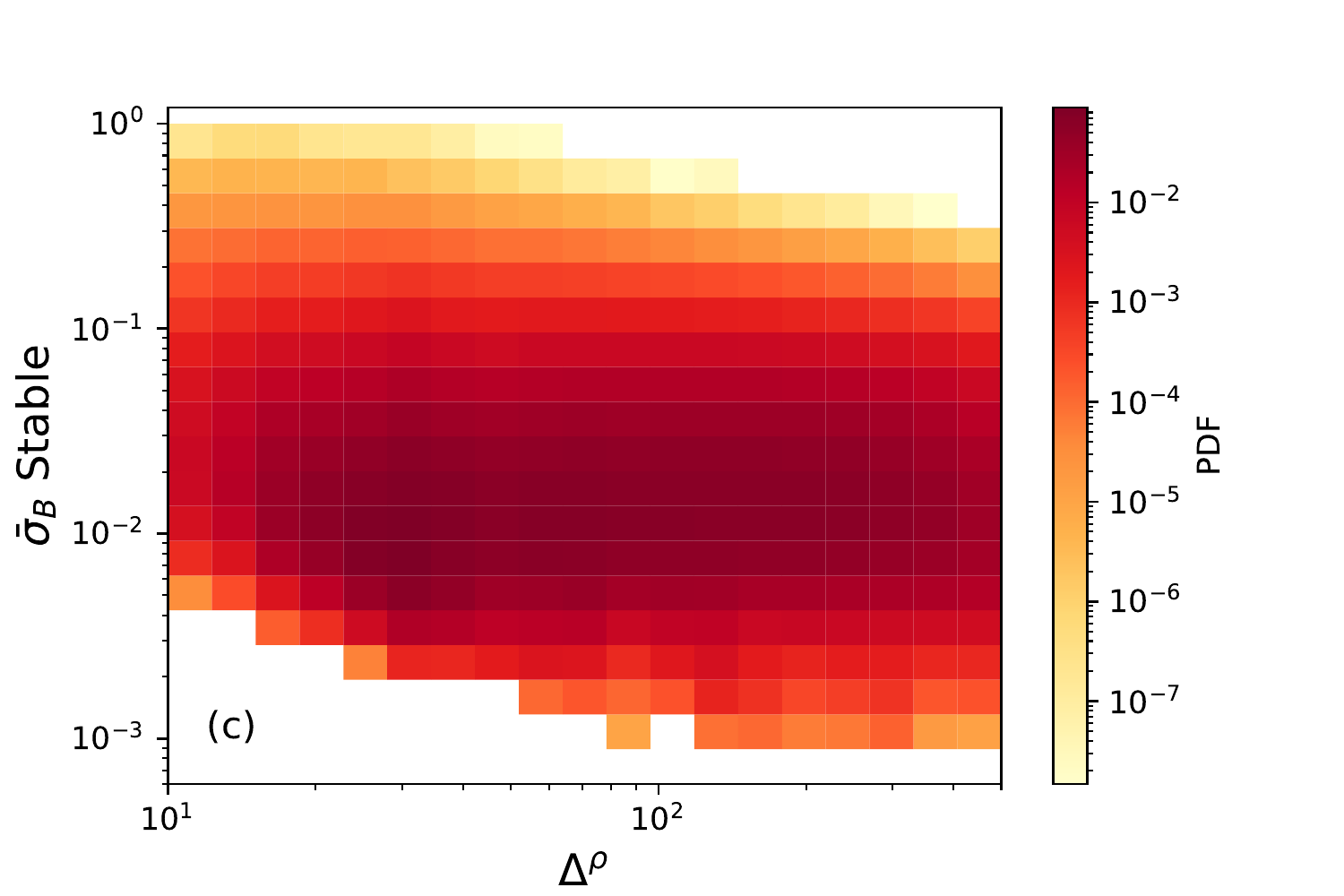}
\caption{PDF of data in $\bar{\sigma}_B$--$\Delta^{\rho}$ parameter space for (a) oblique firehose unstable and (b) mirror-mode unstable data distributions. Panel (c) shows the same PDF for equivalent persistence intervals sampled from the stable data. The vertical lines shown in (a) and (b) denote the breakpoints previously identified by \citet{opie_conditions_2022}. 
\label{fig3draft}}
   
  \end{figure}

%
  

\subsection{Definition of Probability Density Function (PDF)}

We define a datapoint as ``unstable'' if it lies above the threshold given by Eq.~(\ref{ethresh}) for the given instability.
We emphasise that the presence of data in the regions unstable to the oblique firehose and mirror-mode instabilities is a rare occurrence in our overall dataset, representing $\sim 3\%$ and $\sim0.5\%$, respectively, of the total dataset. Consequently, we define the probability density function (PDF) as the normalised density bin count for each individual dataset (i.e., separated by oblique firehose unstable (OF), mirror-mode unstable (M), and stable (S)): 
    \begin{equation} \label{eno}
        \mathrm{PDF}(k) = \frac{\psi_{Ik}}{\psi_{Sk} W_{bk}},
    \end{equation}
where $k \in [\mathrm{OF}, \mathrm{M}, \mathrm S]$, $\psi_{Ik}$ is the raw individual bin count of datapoints in dataset $k$, $\psi_{Sk}$ is the total bin count of dataset $k$ summed across all bins, and $W_{bk}$ is the bin width.
In using Eq.~(\ref{eno}), the distributions for each individual dataset are normalised so that $\sum(W_{bk}\,\mathrm{PDF}(k)) = 1$ for each $k \in [\mathrm{OF}, \mathrm{M}, \mathrm S]$.

\section{Results}
\label{Res}

\subsection{$\sigma_B$ and its distribution in $T_{\perp}/T_{\parallel}$--$\beta_{\parallel}$ parameter space}

In Figure~\ref{fig1draft}(a), we show the binned distribution of datapoints  in the $T_{\perp}/T_{\parallel}$--$\beta_{\parallel}$ parameter space. Each bin is colour-coded with its average value $(\sum\sigma_B)/\psi_I$ of $\sigma_B$ on a logarithmic scale. For the instability thresholds, we use Eq.~(\ref{ethresh}) with fit parameters for a maximum growth rate of  $\gamma_m  = 10^{-2} \Omega_p$, where $\Omega_p$ is the proton gyrofrequency, given by \citet{verscharen_collisionless_2016}. Higher values of $\sigma_B$ occur in the stable data distribution approaching the instability thresholds. In the regions above the thresholds, which overall constrain the data distribution, we see the highest values of the averaged $\sigma_B$.

In Figure~\ref{fig1draft}(b), we show the PDF according to Eq.~(\ref{eno}) of $\sigma_B$ for data defined as stable or unstable to either the oblique firehose or mirror-mode instability. We plot the ensemble mean values of $\sigma_B$ for each of the three distributions as vertical lines. The lowest observed values of $\sigma_B$ for the unstable data are higher than for the stable data. The PDFs for the unstable datasets are biased towards higher values of $\sigma_B$ relative to the PDF for the stable dataset. We find that $[\sigma_{BM}]$ is greater than $[\sigma_{BF}]$, and $[\sigma_{BF}]$ is greater than $[\sigma_{BS}]$, where $[\cdot]$ is the ensemble mean.

\subsection{$\sigma_B$ and its relation to $\theta_{BV}$ parameter space}

In Figure~\ref{fig2draft}, we show PDFs of our data in $\sigma_B$--$\theta_{BV}$ parameter space, where  $\theta_{BV}$ is the angle between the magnetic field $\Vec{B}$ and the solar wind proton bulk velocity $\Vec{V}$ for each measurement point, given as a value between $0^{\circ}$ and $360^{\circ}$ measured clockwise from $\Vec{V}$ in the $\Vec{V} - \Vec{B}$ plane when looking down on the $\Vec{V} - \Vec{B}$ plane from the north. To obtain $\theta_{BV}$, we first calculate the cone angle between $\vec B$ and $\vec V$ using the complete 3D vectors in RTN coordinates as 
\begin{equation} \label{eang}
    \theta_{BV}^{\prime} = \text{arccos}\frac{\vec B\cdot \vec V}{BV}, 
\end{equation}
where $\vec V$ is the bulk velocity of the protons. 
We define the complex numbers $b\equiv B_R+iB_T$ and $v\equiv V_R+iV_T$. We then calculate the angle  $\phi_v=\mathrm{arg}(v)$, where $\mathrm{arg}(\cdot) \in [0,2\pi)$ is the polar angle in the complex plane. After rotating $b$ by $-\phi_v$ in the complex plane, we define the difference angle between $b$ and $v$ as $\phi_{bv}=180^{\circ}\,\mathrm{arg}(be^{-i\phi_v})/\pi$. If $0< \phi_{bv}\le 180^{\circ}$, we set $\theta_{BV}=360^{\circ}-\theta_{BV}^{\prime}$. Otherwise,  we set $\theta_{BV}=\theta_{BV}^{\prime}$ \citep{opie_conditions_2022}.

The distribution of data identified as oblique firehose unstable in Figure~\ref{fig2draft}(a) is clustered around values of $\theta_{BV} \approx 75^{\circ}$ and $\theta_{BV} \approx 255^{\circ}$, which represents a quasi-perpendicular alignment  between $\Vec{B}$ and $\Vec{V}$ \citep[which is consistent with the geometry found by][]{opie_conditions_2022}. 
The distribution of  data identified as mirror-mode unstable in Figure~\ref{fig2draft}(b) exhibits four clusters at $\theta_{BV} \approx 20^{\circ}, 160^{\circ}, 220^{\circ},$ and $340^{\circ}$, which represent a quasi-parallel or quasi-anti-parallel alignment between $\Vec{B}$ and $\Vec{V}$ \citep[see also][]{opie_conditions_2022}.  
The distribution of stable data in Figure~\ref{fig2draft}(c) assumes its maximum values in the range  $0.001\lesssim\sigma_B\lesssim 0.1$, largely independent of $\theta_{BV}$.

\subsection{$\sigma_B$ and its relation to the persistence of unstable intervals}

Figure~\ref{fig3draft} panels (a) and (b) show PDFs  according to Eq.~(\ref{eno}) of data in $\bar{\sigma}_B$--$\Delta^{\rho}$ parameter space, where $\Delta^{\rho}$ is the spatial persistence of consecutive unstable 4\,s intervals in units of the proton gyroradius. As discussed by \citet{opie_conditions_2022}, we calculate $\Delta^{\rho}$ using Taylor's hypothesis \citep{taylor_spectrum_1938}. We identify an interval, $i$, with each unstable datapoint in the dataset for both oblique firehose and mirror-mode instabilities. We calculate the lengthscale $l_i=V_i\tau$ for each unstable interval $i$, where $V_i$ is the proton bulk velocity of interval $i$ and $\tau=4\,\mathrm s$ is the PAS sampling cadence. Using the proton gyroradius $\rho_{\mathrm pi}$ for each individual interval $i$, we then calculate the dimensionless lengthscale $\delta_i^\rho=l_i/\rho_{\mathrm pi}$. We then define
\begin{equation}\label{Delta_rho}
    \Delta_j^\rho=\sum\limits_i \delta_i^\rho
\end{equation}
as the normalised persistence interval for each occurrence of the respective instability as measured at the spacecraft.
 
We define the average $\sigma_B$ over consecutive unstable 4\,s intervals as
\begin{equation}
\bar{\sigma}_B = \frac{1}{n}\sum\limits _{i=1}^{n}\sigma_B\left(t_i\right),
\end{equation}
where $n$ is the number of temporally consecutive 4\,s intervals $t_i$ in each unstable persistence interval of size $\Delta ^{\rho}$. We then bin the data in two-dimensional histograms in $\bar{\sigma}_B$--$\Delta^{\rho}$ space. We show plots for both oblique firehose (Figure~\ref{fig3draft}(a)) and mirror-mode (Figure~\ref{fig3draft}(b)) unstable data.

In Figure~\ref{fig3draft}(c), we show a similar plot for consecutive intervals sampled from the stable dataset. We select all intervals of $P$ consecutive points where $P\in[2,3,4,\dots,14,15]$ and calculate $\bar{\sigma}_B$ and $\Delta^{\rho}$ for the stable data intervals in the same way as for the unstable persistence intervals.

\citet{opie_conditions_2022} identify the breakpoints $\Delta^{\rho}_b$ of the $\Delta^{\rho}$ distribution  as the minimum spatial scales required for these instabilities to act. We overplot $\Delta^{\rho}_b$  as vertical dashed lines in Figure~\ref{fig3draft}(a) and (b). For both unstable modes, the $\bar{\sigma}_B$ value associated with the maximum of the PDF decreases with increasing  $\Delta^{\rho}$. The distributions exhibit  a lower bound at $\bar{\sigma}_B \approx 3\times10^{-3}$ for the oblique firehose and at $\bar{\sigma}_B \approx 4\times10^{-3}$ for the mirror-mode instability. The maximum of the PDF lies near this lower bound at $\Delta^{\rho}\approx\Delta^{\rho}_b$ for each of the unstable modes.

\section{Discussion and interpretation}

\subsection{Distributions in parameter space}

Figure~\ref{fig1draft}(a) shows a clear dependence of $\sigma_B$ on $\beta_{\parallel}$, consistent with previous results using  $|\delta{\Vec{B}}|/B_0$ instead of $\sigma_B$, where $B_0$ is the averaged background magnetic field   \citep{kasper_windswe_2002,bale_magnetic_2009,servidio_proton_2014}. Higher values of $\beta_{\parallel}$ often imply lower values of $B_0$ due to their explicit interdependence in Eq.~(\ref{beta}). This interdependence creates a correlation between $\delta \vec B/B_0$ and $\beta_{\parallel}$ even if $\delta \vec B$ is constant, which is consistent with the overall $\beta_{\parallel}$ dependence of $\sigma_B$ in Figure~\ref{fig1draft}(a). In our analysis, we take this dependence as an inherent feature of the $T_{\perp}/T_{\parallel}$--$\beta_{\parallel}$ parameter space and focus on the observed values of $\sigma_B$ relating to the partition of the space between stable and unstable data.

The joint dependency of the data distributions on $\sigma_B$ and $\theta_{BV}$ shown in Figure~\ref{fig2draft} is consistent with our previous work that shows that the $\theta_{BV}$-dependent anisotropy is opposite to the expectations from adiabatic expansion alone \citep{opie_conditions_2022}. The observed distributions are also consistent with the PDFs in Figure~\ref{fig1draft}(b) which show that the distributions of $\sigma_B$ are skewed towards higher values for data in the unstable parameter regimes and have a higher ensemble mean compared with the stable data distribution. These statistical properties indicate that the relative level of fluctuations on the 4\,s scale, whether from instabilities or background turbulence, is greater in the regions of parameter space unstable to the oblique firehose and mirror-mode instabilities than in the stable regime. The conjunction between Figure~\ref{fig2draft} and our previous work \citep{opie_conditions_2022} points towards a potential role for the fluctuations represented by $\sigma_B$ in raising the tangential and normal temperatures $T_{T}$ and $T_{N}$ relative to the radial temperature $T_R$. We postpone a more detailed discussion of this aspect to future work.

\subsection{Instabilities in a turbulent background}

Our $\sigma_B$ measure captures non-compressive fluctuations at a 4\,s timescale by calculating the full directional variability of the magnetic field. $\sigma_B$ includes fluctuations both from the background turbulence and from the instabilities, as long as the fluctuations have a directional component (e.g.,  Alfv\'enic). Previous work interprets an enhanced level of small-scale fluctuations ($|\delta\Vec{B}|/B_0$) at and beyond the instability thresholds as  evidence of the growing fluctuations of the instabilities \citep{bale_magnetic_2009}. Comparing our Figure~\ref{fig1draft}(a) with the second panel of Figure 1 by \citet{bale_magnetic_2009}, we find that both measures agree quite closely for the oblique firehose instability. In the case of the mirror-mode instability, however, our measure identifies a lower level of enhanced fluctuations than the measure used by \citet{bale_magnetic_2009}, particularly at lower $\beta_\parallel$. We attribute this difference to the predominantly compressive polarisation  of the mirror-mode instability that we intentionally do not capture. We infer that the fluctuations measured by $\sigma_B$ include a significant contribution from background turbulence.  

We make the assumption that turbulent fluctuations create non-equilibrium features \citep{marsch_turbulence_1991,matteini_parallel_2006,schekochihin_nonlinear_2008,maruca_what_2011,matteini_ion_2012}, while instabilities -- once triggered and effective -- reduce non-equilibrium features \citep{gary_mirror_1992,gary_proton_2001,kasper_windswe_2002,hellinger_solar_2006,bale_magnetic_2009}. This assumption suggests that the observed persistence of data in the regions of unstable $T_{\perp}/T_{\parallel}$--$\beta_{\parallel}$ parameter space is evidence that (a) there is insufficient spatial scale for the instabilities to act effectively \citep{opie_conditions_2022}, or (b) that the instabilities cannot immediately overcome the turbulent driving of anisotropy \citep{osman_proton_2013}. A combination of both cases is possible. 

In the ongoing competition between the turbulent driving and the instabilities, the relevant timescales for the opposing processes are important for deciding the outcome. Under stable solar wind conditions, non-linear processes are effective on timescales that are shorter than the linear timescales associated with the instabilities \citep{matthaeus_nonlinear_2014,klein_majority_2018}. However, in the unstable regions of the $T_{\perp}/T_{\parallel}$--$\beta_{\parallel}$ parameter space, the plasma assumes conditions in which the linear timescales associated with the instabilities are equivalent to or shorter than the non-linear timescales associated with the turbulent driving \citep{bandyopadhyay_interplay_2022}. This inversion of the relevant timescales allows the instabilities to provide an effective boundary to non-equilibrium conditions in the solar wind.

\subsection{The interactions between instabilities and turbulence}

If the observed fluctuations measured by $\sigma_B$ were ergodic, which we define as $\langle{\sigma_B}\rangle~=~[\sigma_B]$, we would not expect $\bar{\sigma}_B$ to exhibit dependency on $\Delta^{\rho}$ \citep{matthaeus_stationarity_1982}. The reason for this expectation is that the time-averaged amplitude of the fluctuations at the 4\,s scale, if the fluctuations were ergodic, would not depend on the persistence length $\Delta^{\rho}$ of the intervals over which $\sigma_B$ is averaged\footnote{In our definition of ergodicity, we rely on the assumption, common to other studies, \citep[e.g.,][]{hellinger_solar_2006,bale_magnetic_2009}, that the size of our complete dataset is sufficient to be representative of the statistical properties of solar-wind processes, irrespective of the actual sample size. At the spatial scales we consider here, this assumption is justified \citep{matthaeus_stationarity_1982}.}. For the stable dataset, the distribution of $\bar{\sigma}_B$ does not depend on the averaging length, as shown in Figure~\ref{fig3draft}(c). We verify that $\langle{\sigma_{BP}}\rangle\approx [\sigma_B] \approx~0.032$, where $\langle{\sigma_{BP}}\rangle$ is the mean value of $\sigma_B$ for stable intervals of length $P$ and $[\sigma_B]$ is the ensemble mean for the complete dataset of stable datapoints, taken as representative of the statistical properties of the stable solar wind. Subject to our definition, the condition $\langle{\sigma_{BP}}\rangle\approx [\sigma_B]$ indicates ergodicity. However, in Figure~\ref{fig3draft} (a) and (b), the distribution of the data in $\bar{\sigma}_B$--$\Delta^{\rho}$ parameter space indicates an interdependency between  $\bar{\sigma}_B$ and $\Delta^{\rho}$ for both oblique firehose and mirror-mode unstable data. This interdependency suggests that $\sigma_B$ is not ergodic for the unstable intervals and therefore that the unstable intervals are statistically disjoint from the stable intervals \citep{matthaeus_stationarity_1982,walters_introduction_2000}. We infer that the interdependency is indicative of processes that are only relevant to the unstable regimes. From our previous assumption, these processes relate either to the creation of non-equilibrium features by background turbulence or to the action of instabilities to reduce non-equilibrium features. In both cases, the process concerned must disrupt the ergodicity of the turbulent fluctuations measured by $\sigma_B$ for the stable regime.

The distributions in Figure~\ref{fig3draft}(a) and (b) show that unstable intervals are more likely to be larger in units of $\Delta^{\rho}$ when $\bar{\sigma}_B$ is lower. The highest probability densities of the distribution of unstable data are observed and maintained for values of $\Delta^{\rho}<\Delta^{\rho}_b$, which we identify as the persistence intervals in which instabilities do not act effectively \citep{opie_conditions_2022}. In these intervals, higher $\bar{\sigma}_B$ implies shorter residence time for the plasma in any particular unstable regime of $T_{\perp}/T_{\parallel}$--$\beta_{\parallel}$ parameter space, largely independently of the action of instabilities. 

The interdependency continues when $\Delta^{\rho}>\Delta^{\rho}_b$, which we identify as the persistence intervals in which instabilities do act effectively \citep{opie_conditions_2022}. 
For these intervals, Figure~\ref{fig3draft} shows that longer unstable intervals (in terms of $\Delta^{\rho}$) are more likely to have a lower value of $\bar{\sigma}_B$ than shorter unstable intervals. 

We interpret the value of $\bar{\sigma}_B$ as a measure for turbulent ``activity''. Likewise, we interpret a lower PDF value for unstable intervals as an indication of the more efficient action of the instabilities. In this interpretation, the observed likelihood trend suggests that the efficiency of instabilities to reduce temperature anisotropy is greater in larger and more active intervals than in shorter and less active intervals. Therefore, the competition between the linear relaxation time and the nonlinear time not only depends on $\bar{\sigma}_B$ (i.e., a measure for the nonlinear time) but also on $\Delta^{\rho}$. 

\subsection{Limitations of our analysis}

In our analysis, we do not include the roles of the parallel firehose or ion-cyclotron instabilities. In general,  the non-propagating oblique firehose and mirror-mode instabilities are more effective in constraining temperature anisotropy \citep{gary_theory_1993,gary_proton_1997,kunz_firehose_2014,gary_short-wavelength_2015,rincon_non-linear_2015}. The thresholds for these instabilities are calculated from linear theory under the assumption of conditions that do not exactly apply to the turbulent solar wind \citep{matthaeus_nonlinear_2014}. Nonetheless, observational studies have shown that these thresholds usefully define the boundaries of the stability of the plasma \citep{hellinger_solar_2006,bale_magnetic_2009,gary_short-wavelength_2015,chen_multi-species_2016}. It remains an open question as to why the non-propagating thresholds provide better constraints to the data distribution in $T_{\perp}/T_{\parallel}$--$\beta_{\parallel}$ parameter space even when the propagating instabilities have lower theoretical thresholds \citep{gary_short-wavelength_2015,markovskii_proton_2019,verscharen_multi-scale_2019}.

The directional variations measured by $\sigma_B$ have an impact on the measurement of $T_\perp$ and $T_\parallel$. The relevant timescale for this measurement is the 1\,s SWA/PAS sampling interval. The typical directional variation in $\Vec{B}$ over one second is $\sim 3.4^\circ$ for our dataset and thus small compared to the angular resolution of PAS. However, at large $\sigma_B\gtrsim 0.5$, the deflections are potentially significant. Therefore caution must be exercised when defining the instability of intervals at large $\sigma_B\gtrsim 0.5$.

\section{Conclusions}

We show that non-compressive magnetic field variability, $\sigma_B$, is a useful measure for evaluating the interplay between turbulence and instabilities in the solar wind. Background magnetic field fluctuations cascade to the small-scale end of the inertial range where they have the ability to increase the temperature anisotropy. If the anisotropy is sufficiently large, the plasma becomes unstable. 

The distribution of the data in $\bar{\sigma}_B$--$\Delta^{\rho}$ parameter space shows that $\bar{\sigma}_B$ and $\Delta^{\rho}$ are interdependent only for the unstable plasma intervals. The competition between the action of the turbulence and the instabilities in these unstable intervals depends on both the level of turbulent activity and the spatial persistence of conditions that define the oblique firehose and mirror-mode instabilities. Our analysis suggests that the turbulent solar wind does not provide a simple homogeneous background as assumed by classical linear theory. In fact, a complex interaction between turbulent fluctuations and kinetic instabilities ultimately regulates the proton-scale energetics of the solar wind.

\begin{acknowledgements}

We appreciate very valuable discussions with Matt Kunz. For the purpose of open access, the author has applied a Creative Commons Attribution (CC BY) licence to any Author Accepted Manuscript version arising. 
S.~O.~is supported by the Natural Environment Research Council (NERC) grant  NE/S007229/1. D.~V.~is supported by the Science and Technology Facilities Council (STFC) Ernest Rutherford Fellowship ST/P003826/1. D.~V.~and C.~J.~O. are supported by STFC Consolidated Grants  ST/S000240/1 and ST/W001004/1. C.~H.~K.~C.~is supported by UKRI Future Leaders Fellowship MR/W007657/1 and STFC Consolidated Grant ST/T00018X/1. P.~A.~I.~is supported by NASA grant 80NSSC18K1215 and by National Science Foundation (NSF) grant AGS2005982. Solar Orbiter is a space mission of international collaboration between ESA and NASA, operated by ESA. Solar Orbiter Solar Wind
Analyser (SWA) data are derived from scientific sensors which have been
designed and created, and are operated under funding provided in
numerous contracts from the UK Space Agency (UKSA), STFC, the Agenzia Spaziale Italiana
(ASI), the Centre National d’Etudes Spatiales (CNES), the Centre
National de la Recherche Scientifique (CNRS), the Czech
contribution to the ESA PRODEX programme and NASA. Solar Orbiter SWA
work at UCL/MSSL is currently funded under STFC grants ST/T001356/1 and
ST/S000240/1. The Solar Orbiter magnetometer was funded by UKSA grant ST/T001062/1. This work was discussed at the ISSI Team ``Ion Kinetic Instabilities in the Solar Wind in Light of Parker Solar Probe and Solar Orbiter Observations'', led by L.~Ofman and L.~Jian.

\end{acknowledgements}

%
%

\bibliography{letter1}
\bibliographystyle{aa}

\end{document}